%% file: main.tex
\documentclass[conference]{IEEEtran}
\IEEEoverridecommandlockouts
\usepackage{cite}
\usepackage{amsmath,amssymb,amsfonts}
\usepackage{algorithmic}
\usepackage{graphicx}
\usepackage{textcomp}
\usepackage{xcolor}
\usepackage{float}
\usepackage{multirow}
\usepackage[none]{hyphenat}
\usepackage{tikz}

\newcommand{\copyrighttext}{%
  \footnotesize \textcopyright 2021 IEEE. DOI: 10.1109/WAMICON47156.2021.9443612 Personal use of this material is permitted.
  Permission from IEEE must be obtained for all other uses, in any current or future
  media, including reprinting/republishing this material for advertising or promotional
  purposes, creating new collective works, for resale or redistribution to servers or
  lists, or reuse of any copyrighted component of this work in other works. 
}
\newcommand{\copyrightnotice}{%
\begin{tikzpicture}[remember picture,overlay]
\node[anchor=south,yshift=10pt] at (current page.south) {\fbox{\parbox{\dimexpr\textwidth-\fboxsep-\fboxrule\relax}{\copyrighttext}}};
\end{tikzpicture}%
}

\def\BibTeX{{\rm B\kern-.05em{\sc i\kern-.025em b}\kern-.08em
    T\kern-.1667em\lower.7ex\hbox{E}\kern-.125emX}}

\input{IMS_modify_IEEEtran_18b_CTAN_V4c}

\begin{document}

\title{Plasma Switch-Based Technology for High-Speed and High-Power
Impedance Tuning
\thanks{This effort undertaken was sponsored by the Department of the Navy, Office of Naval Research under ONR award number N00014-19-1-2549. This work relates to Department of Navy award N00014-19-1-2549 issued by the Office of Naval Research. The United States Government has a royalty-free license throughout the world in all copyrightable material contained herein. Any opinions, findings, and conclusions or recommendations expressed in this material are those of the author(s) and do not necessarily reflect the views of the Office of Naval Research.}
}
\IMSauthor{%
\IMSauthorblockNAME{
Zach Vander~Missen\IMSauthorrefmark{\#1},
Sergey Macheret\IMSauthorrefmark{\$2}
Abbas Semnani\IMSauthorrefmark{*3},
Dimitrios Peroulis\IMSauthorrefmark{\#4}
}
\\%
\IMSauthorblockAFFIL{
\IMSauthorrefmark{\#}School of Electrical and Computer Engineering, Purdue University, USA\\
\IMSauthorrefmark{\$}School of Aeronautics and Astronautics, Purdue University, USA\\
\IMSauthorrefmark{*}Department of Electrical Engineering and Computer Science, The University of Toledo, USA
}
\\%
\IMSauthorblockEMAIL{
\IMSauthorrefmark{1}zvm@purdue.edu,
\IMSauthorrefmark{2}macheret@purdue.edu,
\IMSauthorrefmark{3}abbas.semnani@utoledo.edu,
\IMSauthorrefmark{4}dperouli@purdue.edu
}
}

%
\maketitle
\copyrightnotice
%
%
%
\begin{abstract}
This paper introduces a new technology for a high-speed, high-power mobile form-factor tuner utilizing gas discharge tube plasma cells as switching components. To the best of our knowledge, this represents the first plasma-enabled RF matching network. Technology development is reviewed, the fabrication and measurement of a proof-of-concept switched stub impedance tuner are presented, and techniques for improvement are discussed. The proof-of-concept impedance tuner functions with a 27\% bandwidth from 3 to almost 4~GHz and shows a power gain better than -2.5~dB across all switching state\nobreakdash-frequency combinations at a 50~W input power level with spread coverage of the Smith chart. State change transient timing is measured to be on the order of 500~ns. This technology demonstration highlights the potential of miniaturized, rapidly-tunable, high-power, plasma-based RF devices.
\end{abstract}
\begin{IEEEkeywords}
Gas discharge tube (GDT), high-power, impedance tuner, plasma, plasma switch.
\end{IEEEkeywords}
%
%

\section{Introduction}

Electronically controllable impedance tuners are an important category of devices with wide-ranging applications. They find use in load- and source-pull, re-configurable amplifiers, antenna array matching, and beam-forming, among others. One widely-used solution for impedance tuning is the mechanical impedance tuner. These devices can be high-power tolerant and exhibit excellent RF performance, but they are typically large, heavy, and slow. They offer a good solution for lab characterization, but for dynamic environments like radar and electronic warfare applications, they are not practical \cite{IMS2020:PC2014,IMS2020:MM}. On the other end of the spectrum, there are impedance tuning solutions based on semiconductor devices like varactor diodes, MEMS-based solutions, tunable cavity resonators, and ferroelectric capacitors. These solutions exhibit complementary issues to mechanical tuners in that some can be much faster, but most suffer from their own drawbacks such as non-linearity and low power handling \cite{IMS2020:JF2008,IMS2020:RW2006,IMS2020:YL2005,IMS2020AS2017T,IMS2020:AS2019J}. \\
One conventional topology paired with many of these technologies is the switched stub impedance tuner. Semiconductor, MEMS, or other switch technologies are controlled to selectively connect or disconnect several transmission line stubs to a main transmission line. Through this reconfiguration, input impedance tuning of the device is achieved. However, as power levels increase, the availability of quality switching elements decreases. Semiconductor-based switches suffer from nonliterary, and MEMS switches are unable to operate at increasingly high power levels.\\ Plasma-based designs offer a compelling solution to fill this apparent gap in the field. Plasma devices have been demonstrated in various topologies, \cite{IMS2020:AS2016,IMS2020:AS2016J,IMS2020:AS2018J,IMS2020:ZVM2019}, demonstrating again and again benefits in these areas of need, namely speed, power handling, and linearity \cite{IMS2020:AS2018J,IMS2020:ZVM2019}. Developing further the existing work evaluating plasma switching, \cite{IMS2020:AS2018J,IMS2020:AS2017}, this work demonstrates that the benefits of plasma as a technology can be brought to bear for demanding impedance tuning applications.

\begin{figure}
\centering
\includegraphics[width=2.7 in]{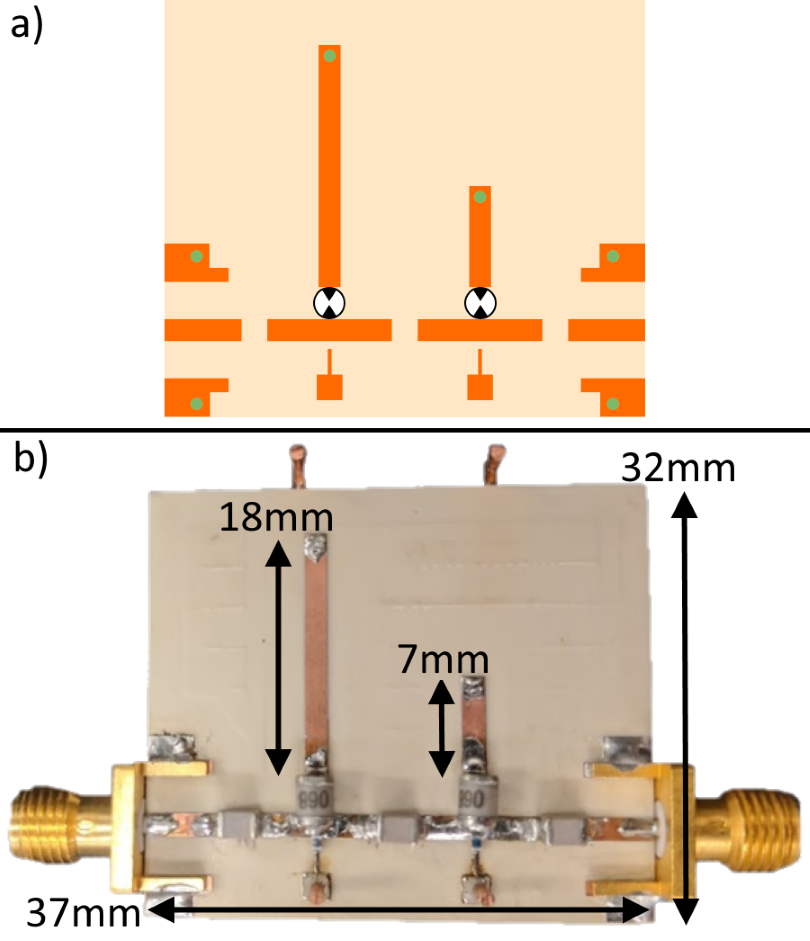}
\caption{a) Schematic layout for the proof-of-concept impedance tuner with plasma cell locations indicated with the GDT schematic symbol. b) The assembled proof-of-concept impedance tuner with dimensions indicating the outer board dimensions and tuning stub lengths.}
\label{fig:board}
\end{figure}


{
%
\setlength{\tabcolsep}{1.25mm}%
\renewcommand{\arraystretch}{1.2}
\newcommand{\CPcolumnonewidth}{not used}%
\newcommand{\CPcolumntwowidth}{23mm}%
\newcommand{\CPcolumnthreewidth}{12mm}%
\newcommand{\CPcolumnfourwidth}{33mm}%
\begin{table}[hbp]
\caption{Components used in proof-of-concept impedance tuner. Capacitors C4 and C5 are bottom-mounted and not visible in Fig. \ref{fig:board}.}
\small
\centering
\begin{tabular}{|l|l|l|l|}\hline
{\bfseries Component} & {\bfseries Value} & {\bfseries Manufacturer} & {\bfseries Model number} \\ \hline
C1, C2, C3 & 7.5~pF & Johanson & 102S42E7R5CV3E \\ \hline
C4, C5 & 7.5~pF & Johanson & 501S42E7R5CV4E \\ \hline
L1, L2 & 15~nH & Coilcraft & 0403HQ-15N \\ \hline
GDT1, GDT2 & 200~V & Littelfuse & CG7200MS \\ \hline
J1, J2 & SMA & Pasternack & PE4543 \\ \hline
\end{tabular}
\label{tab:components}
\end{table}
}

\section{Technology Development}
\subsection{Approach}
\begin{figure}[bp!]
\centering
\includegraphics[width=3.25 in]{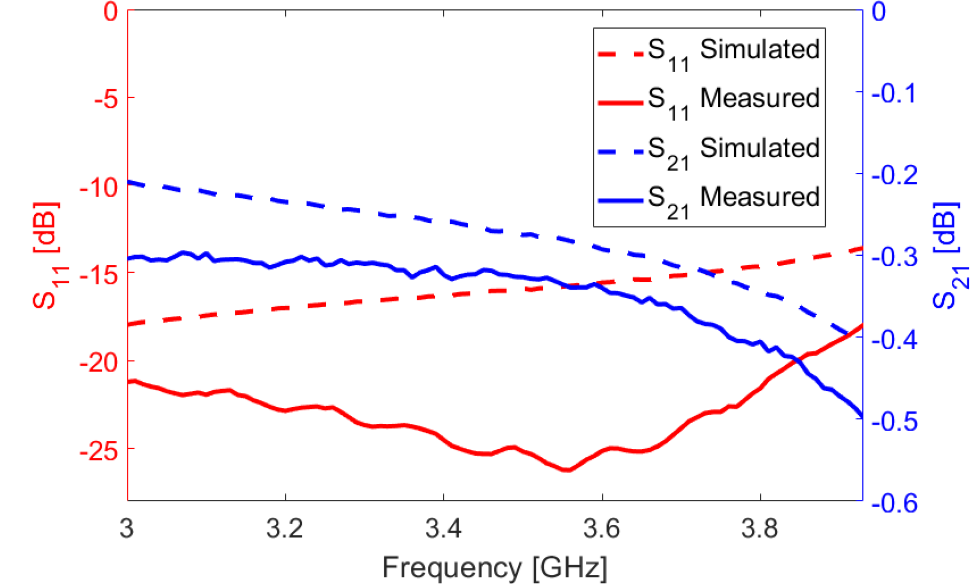}
\caption{Measured and simulated S-parameters of the proof-of-concept impedance tuner in the 00 state (no stubs switched into the circuit).}
\label{fig:sim2sp}
\end{figure}

\begin{figure}[t]
\centering
\includegraphics[width=3.25 in]{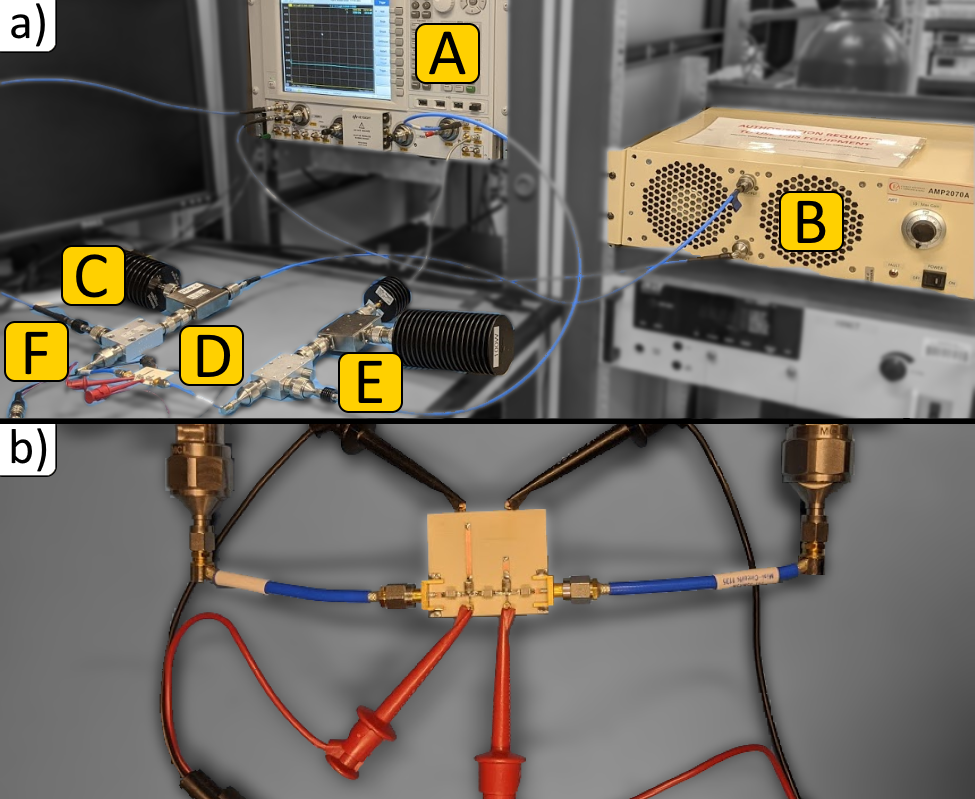}
\caption{a) The measurement setup for high-power S-Parameter measurement A) N5247A VNA B) high-power amplifier C) port 1 external, modified, high-power test set D) device under test E) port 2 external, modified, high-power test set F) GDT switch control bias lines. b) Close-up of proof-of-concept impedance tuner connected to the high-power test set.}
\label{fig:measurement}
\end{figure}

The permittivity and conductivity of a plasma can be tuned by varying its electron number density ($n_e$) \cite{IMS2020:RZ}. This behavior is described by (\ref{eqn:ep}) and (\ref{eqn:sig}), where $e$ and $m$ represent the electron's charge and mass, $\epsilon_0$ is the permittivity of free space, $\nu_m$ is the electron-neutral collision frequency, and $\omega$ is the stimulus frequency. In addition to the bulk plasma permittivity and conductivity, the plasma discharge equivalent circuit also includes the capacitance of near-electrode sheaths \cite{IMS2020:RZ}.
\begin{equation}
\epsilon_r = 1-\frac{e^2n_e}{\epsilon_0m(\omega^2+\nu_m^2)},
\label{eqn:ep}
\end{equation}
\begin{equation}
\sigma = \frac{e^2n_e\nu_m}{m(\omega^2+\nu_m^2)} 
\label{eqn:sig}
\end{equation}
In the case of plasma switching, the on and off states can be considered separately. In the off-state, $n_e$ is several orders of magnitude smaller than in the on-state, so $\epsilon_r~\approx~1$ and $\sigma~\approx~0$. The switch's isolation is governed by the capacitance between the plasma electrodes. In the on-state, it is desirable for $n_e$ to be as high as possible, so that $\sigma$ is maximized and the on-state resistance is minimized.\\
In this work, commercially available gas discharge tube (GDT) surge arresters are utilized as plasma cells. GDTs are convenient, since they are widely available, however, they are not specifically optimized for this application. Further optimization to decrease the off-state capacitance and on-state resistance can serve to increase the bandwidth and Smith chart coverage as well as decrease loss.\\
As a demonstration of the value plasma technology brings to impedance tuning, a proof-of-concept switched stub impedance tuner was fabricated. ADS was utilized to design the tuner through detailed schematic simulation. The GDT plasma cells utilized in the tuner design were first independently characterized under high incident RF power using the same measurement setup detailed in Fig. \ref{fig:measurement}. The on and off state behavior of the plasma cells was then represented by data-based models in the simulation environment. The specific chosen cells were selected for their superior low off-state capacitance and withstanding voltage which was sufficient for the targeted 50~W incident RF power level.\\
Fig. \ref{fig:sim2sp} shows the measured and simulated S-parameters of the tuner in its 00 state (no stubs switched into the circuit) and highlights the good agreement between simulation and fabricated results. 

\begin{figure}
\centering
\includegraphics[width=2.637 in]{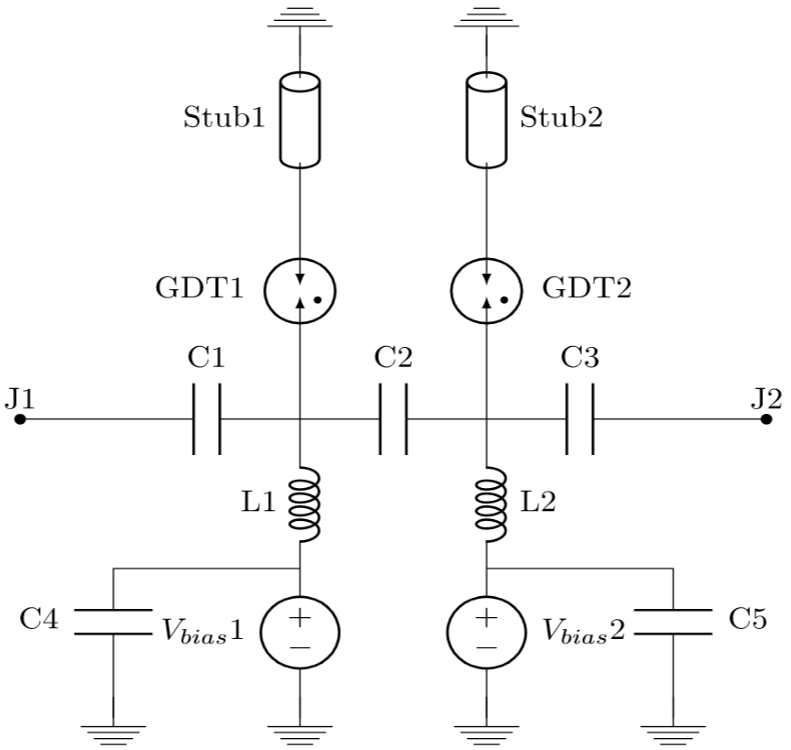}
\caption{A simplified schematic diagram (interconnect transmission lines omitted for clarity) of the proof-of-concept impedance tuner. Labeled component values and models can be found in Table \ref{tab:components}.}
\label{fig:schematic}
\end{figure}

\subsection{Fabrication}

A topology of terminated stubs switched onto a main transmission line was adopted. Short-terminated stubs were selected because they provided improved bandwidth with regard to acceptable power gain in simulation. All additional components depicted in Fig. \ref{fig:schematic} were necessary to provide the required DC biasing to enable switching control of the plasma cells. The application of DC bias across the plasma cells in this design was chosen to guarantee plasma formation in the desired cells. To accomplish this goal, a bias level of approximately 110\% of the nominal rated breakdown voltage of the plasma cell was chosen. It is through this biasing to breakdown that switching control of the plasma cells is achieved. It is important to note that this biasing scheme was implemented using current-limited sources set to limit the bias current to 100~\(\mu\)A. This significant current limit serves to prevent the DC discharge current from prematurely aging the plasma cells and reducing their lifetimes.\\
The proof-of-concept impedance tuner was fabricated on Rogers TMM3 substrate (\(\epsilon_r\)~=~3.45 and tan\(\delta\)~=~0.0020) with a substrate thickness of 30~mil and 17.5~\(\mu\)m copper cladding. Littelfuse CG7200MS GDTs were utilized as plasma cells for the plasma switching components. The device was connectorized using SMA connectors. All capacitors were assembled in vertical orientation to increase the SRF-free bandwidth. A detailed listing of components' model numbers and values corresponding to components in the schematic diagram is available in Table \ref{tab:components}.

{
\setlength{\tabcolsep}{1mm}%
\newcommand{\CPcolumnonewidth}{50mm}%
\newcommand{\CPcolumntwowidth}{91mm}%
\newcommand{\CPcell}[1]{\hspace{0mm}\rule[-0.3em]{0mm}{1.3em}#1}%
\newcommand{\CPcellbox}[1]{\parbox{90mm}{\hspace{0mm}\rule[-0.3em]{0mm}{1.3em}#1\strut}}%
\begin{table*}[htbp!]
\caption{Comparison to state-of-the-art. $^*$Per 10~degrees at 2~GHz. $^\dagger$For 10\% actuator travel.}
\small
\centering
\begin{tabular}{|l|l|l|l|l|l|}\hline
{\bfseries Reference} & {\bfseries Tuning Technology} & {\bfseries Frequency} & {\bfseries Power handling} & {\bfseries Tuning time} & {\bfseries Loss} \\ \hline
\cite{IMS2020:MM} & Mechanical & 0.4-8~GHz & 54~dBm & 550~ms$^*$ & 0.3~dB \\ \hline
\cite{IMS2020:AS2019J} & Linear Actuator & 2.7-4.4~GHz & 50~dBm & 9.2~ms$^\dagger$ & 0.56~dB \\ \hline
\cite{IMS2020:CS2018} & Varactor & 2-2.5~GHz & 44~dBm & -- & 2~dB \\ \hline
\cite{IMS2020:YL2005} & MEMS & 30~GHz & 36~dBm & -- & $>$3~dB \\ \hline
This Work & Plasma & 3-3.93~GHz & 47~dBm & 550~ns & $<$2.5~dB \\ \hline
\end{tabular}
\label{tab:lit}
\end{table*}
}
\begin{figure}[hbp]
\centering
\includegraphics[width=3.25 in]{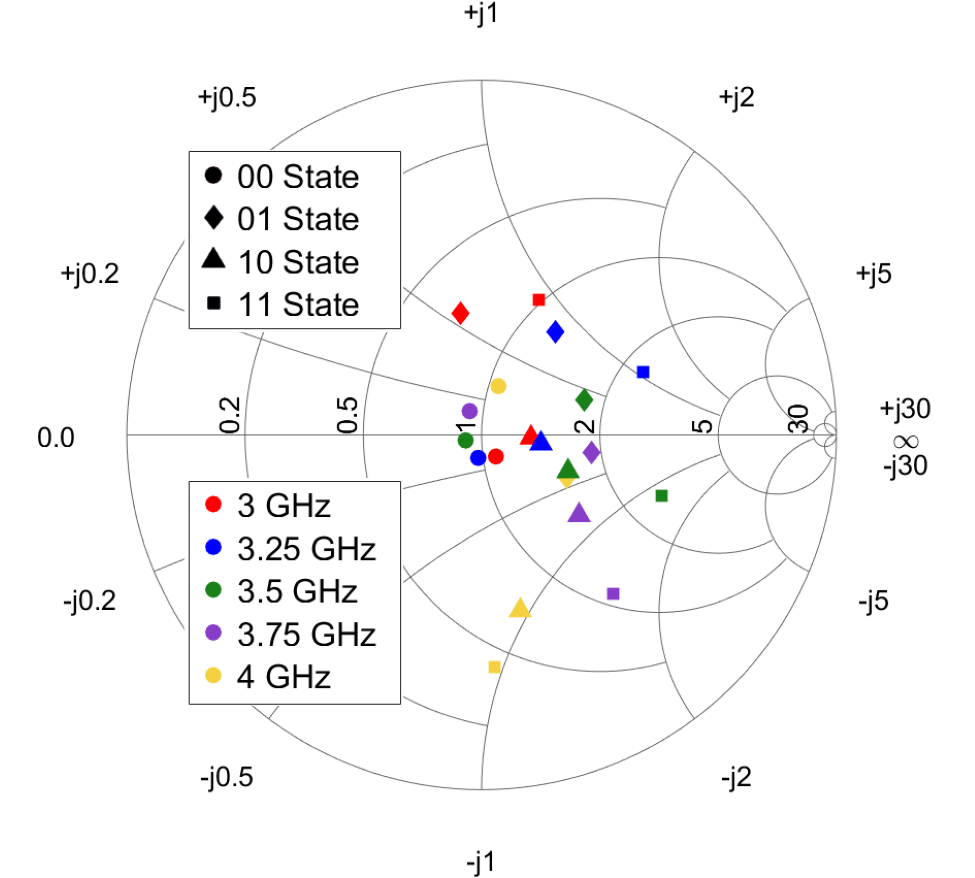}
\caption{Measured Smith chart coverage of the proof-of-concept impedance tuner. Five frequency slices are shown, with points grouped by color. The four switch combination tuning states are indicated by marker shape.}
\label{fig:meas_smith}
\end{figure}
\section{Measurement and Results}

In order to characterize the proof-of-concept impedance tuner, an AMP2070A high-power amplifier was used. The measurement effort was focused on evaluating the impedance tuning performance as well as the power gain of the tuner at RF power levels up to 50~W. To this end, an external, modified test set was utilized with direct receiver access on an N5247A vector network analyzer as depicted in Fig. \ref{fig:measurement}. Source power calibration was used with an external power meter to ensure a level, accurate power at the device input plane. The measurements were half-amplified, meaning that the forward direction stimulus from port 1 was amplified to 50~W, but the reverse direction stimulus from port 2 was at a low, nominal power. In order to calibrate the VNA, the quick SOLT calibration algorithm was utilized as it enabled 2-port calibration without connecting sensitive load standards to the high RF power at port 1.

\begin{figure}[t]
\centering
\includegraphics[width=3.25 in]{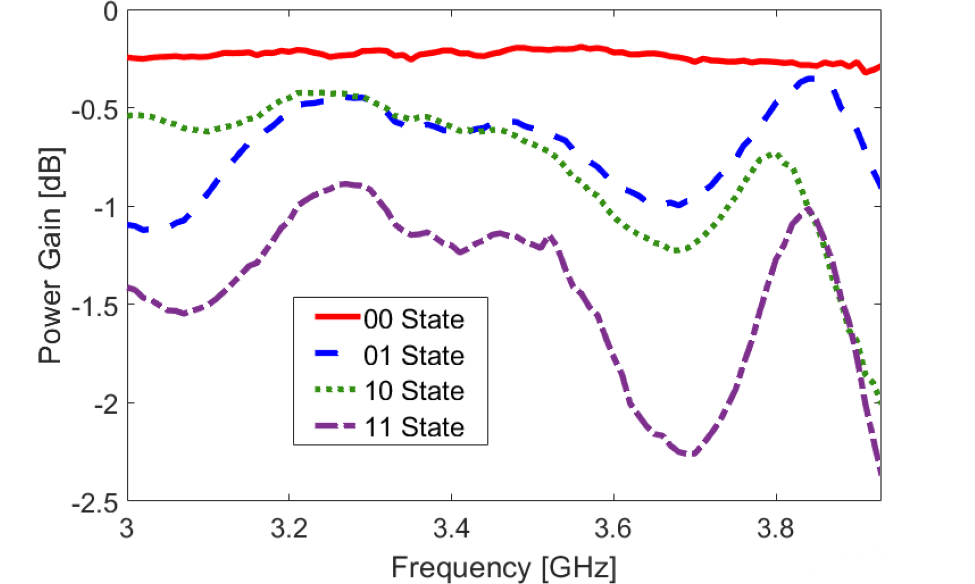}
\caption{Measured power gain of the proof-of-concept impedance tuner for all four switch state combinations with a 50~W input power level. }
\label{fig:meas_tl}
\end{figure}

\begin{equation}
G = \frac{\left|S_{21}\right|^2}{1 - \left|S_{11}\right|^2} 
\label{eqn:tl}
\end{equation}

The power gain (the ratio of the power dissipated in the load to the power delivered to the input of the tuner) of the proof-of-concept tuner was evaluated using the high-power S-parameter measurements. The power gain was found according to (\ref{eqn:tl}) as in \cite{IMS2020:PZ} assuming that the tuner was terminated into a matched load.

As shown in Fig. \ref{fig:meas_tl}, considering a bandwidth from 3-3.93~GHz, the tuner exhibited a worst-case power gain better than -2.5~dB with a 50~W input power level and a power gain better than -1~dB in more than 65\% of all state-frequency combinations. Switching states are identified through a binary code which assigns a zero for an open switch and a one for a closed switch. Switch-stub pairs in order starting closest to port one are encoded from most to least significant bit. 

Fig. \ref{fig:meas_smith} shows the measured $S_{11}$ of the tuner in all four switching states at several frequencies. The proof-of-concept tuner demonstrates a reasonable spread of Smith chart coverage at all frequencies considering the limitation of only 4 tuning states. Extension of the topology to higher stub counts benefits the coverage on the Smith chart significantly, since the state count of an n-stub tuner grows as $2^n$.

\begin{figure}[hb]
\centering
\includegraphics[width=3.25 in]{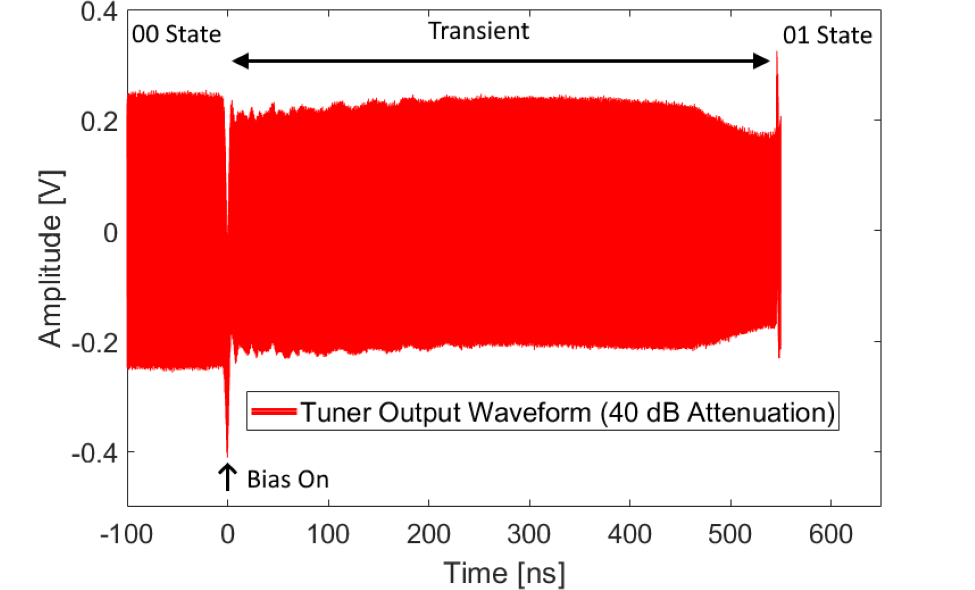}
\caption{A measured state transition transient of the proof-of-concept impedance tuner from the 00 state to the 01 state with a 3~GHz input signal. The plotted trace is the measured output waveform of the tuner with 40~dB of additional attenuation. Time 0 is aligned to the bias on point.}
\label{fig:timing}
\end{figure}

Finding solutions for high-speed impedance tuning at high power levels is a formidable technological challenge. Plasma presents a distinct advantage in this area, since the formation and decay of plasma typically operate at timescales in the nanoseconds \cite{IMS2020:AS2018J}. Fig. \ref{fig:timing} shows the state switching transient of the tuner as it switches from the 00 state to the 01 state with a 3~GHz CW stimulus. This measurement was conducted by measuring the tuner output signal with a high-speed oscilloscope. The control voltage was switched, and the transient waveform was captured. The fabricated tuner shows a tuning time of 550~ns to switch between states. 

As seen in Table \ref{tab:lit}, the performance of the proof-of-concept tuner combines the best of both worlds as compared to state-of-the-art high-power tuners. In combining high power handling and rapid tuning time this plasma technology has the potential to out-preform other solutions.

\section{Conclusion}

A new technology for a high-speed, high-power switched stub impedance tuner utilizing GDT plasma cells as switching components was introduced representing the first ever plasma-based RF matching network demonstrated. Technology development was reviewed, the fabrication and measurement of a proof-of-concept switched stub impedance tuner were presented, and techniques for improvement were discussed. The fabricated impedance tuner functions with a 27\% bandwidth from 3 to almost 4~GHz and shows a power gain better than -2.5~dB across all switching state\nobreakdash-frequency combinations at a 50~W input power level with spread coverage of the Smith chart and state change transient timing measured to be on the order of 500~ns. These results provide strong evidence of the value of plasma switching for high-speed, high-power impedance tuning.


\bibliographystyle{IEEEtran}

\bibliography{IEEEabrv,IMS2020}

\end{document}

%% file: IMS_modify_IEEEtran_18b_CTAN_V4c.tex
\makeatletter

\def\@IMSauthorblockNAMEstyle{\normalfont\IMSauthorsize}
\def\@IMSauthorblockAFFILstyle{\normalfont\IMSaffilsize}
\def\@IMSauthorblockEMAILstyle{\normalfont\IMSaffilsize}
\def\IMSauthorblockNAME#1{%
\relax\@IMSauthorblockNAMEstyle%
#1%
}%
\def\IMSauthorblockAFFIL#1{%
\relax\@IMSauthorblockAFFILstyle%
\vskip\@IEEEauthorblockAtopspace
#1%
}%
\def\IMSauthorblockEMAIL#1{%
\relax\@IMSauthorblockEMAILstyle%
\vskip\@IEEEauthorblockAtopspace
#1%
}%
\newcommand{\IMSauthor}[1]{%
\ifIsBlindReviewVersion%
\author{\phantom{\parbox{\textwidth}{\center\relax#1}}}%
\else%
\author{\parbox{\textwidth}{\center\relax#1}}%
\fi%
}%
\newif\ifIsBlindReviewVersion

\def\@maketitle{\newpage
\bgroup\par\addvspace{0.5\baselineskip}\centering%
\ifCLASSOPTIONtechnote
   {\bfseries\large\@IEEEcompsoconly{\sffamily}\@title\par}\vskip 1.3em{\lineskip .5em\@IEEEcompsoconly{\sffamily}\@author
   \@IEEEspecialpapernotice\par{\@IEEEcompsoconly{\vskip 1.5em\relax
   \@IEEEtitleabstractindextextbox{\@IEEEtitleabstractindextext}\par
   \hfill\@IEEEcompsocdiamondline\hfill\hbox{}\par}}}\relax
\else
   \vskip0.2em{\IMStitlesize\ifCLASSOPTIONtransmag\bfseries\LARGE\fi\@IEEEcompsoconly{\sffamily}\@IEEEcompsocconfonly{\normalfont\normalsize\vskip 2\@IEEEnormalsizeunitybaselineskip
   \bfseries\Large}\@title\par}\vskip1.0em\par
   \ifCLASSOPTIONconference%
      {\@IEEEspecialpapernotice\mbox{}\vskip\@IEEEauthorblockconfadjspace%
       \mbox{}\hfill\begin{@IEEEauthorhalign}\@author\end{@IEEEauthorhalign}\hfill\mbox{}\par}\relax
   \else
      \ifCLASSOPTIONpeerreviewca
         {\@IEEEcompsoconly{\sffamily}\@IEEEspecialpapernotice\mbox{}\vskip\@IEEEauthorblockconfadjspace%
          \mbox{}\hfill\begin{@IEEEauthorhalign}\@author\end{@IEEEauthorhalign}\hfill\mbox{}\par
          {\@IEEEcompsoconly{\vskip 1.5em\relax
           \@IEEEtitleabstractindextextbox{\@IEEEtitleabstractindextext}\par\hfill
           \@IEEEcompsocdiamondline\hfill\hbox{}\par}}}\relax
      \else
         \ifCLASSOPTIONtransmag
           {\@IEEEspecialpapernotice\mbox{}\vskip\@IEEEauthorblockconfadjspace%
            \mbox{}\hfill\begin{@IEEEauthorhalign}\@author\end{@IEEEauthorhalign}\hfill\mbox{}\par
           {\vspace{0.5\baselineskip}\relax\@IEEEtitleabstractindextextbox{\@IEEEtitleabstractindextext}\vspace{-1\baselineskip}\par}}\relax
         \else
           {\lineskip.5em\@IEEEcompsoconly{\sffamily}\sublargesize\@author\@IEEEspecialpapernotice\par
           {\@IEEEcompsoconly{\vskip 1.5em\relax
            \@IEEEtitleabstractindextextbox{\@IEEEtitleabstractindextext}\par\hfill
            \@IEEEcompsocdiamondline\hfill\hbox{}\par}}}\relax
         \fi
      \fi
   \fi
\fi\par\addvspace{0.0\baselineskip}\egroup}

\def\IMStitlesize{\@setfontsize{\IMStitlesize}{18}{21pt}}
\def\IMSauthorsize{\@setfontsize{\IMSauthorsize}{12}{13pt}}
\def\IMSaffilsize{\@setfontsize{\IMSaffilsize}{12}{13pt}}
\def\IMScaptionsize{\@setfontsize{\IMScaptionsize}{8}{9pt}}
\def\IMSbibsize{\@setfontsize{\IMSbibsize}{8}{9pt}}

\def\@IEEEauthorblockNstyle{\IMSauthorsize\@IEEEcompsocnotconfonly{\sffamily}\@IEEEcompsocconfonly{\large}}
\def\@IEEEauthorblockAstyle{\IMSaffilsize\@IEEEcompsocnotconfonly{\sffamily}\@IEEEcompsocconfonly{\itshape}\@IEEEcompsocconfonly{\large}}
\def\@IEEEauthordefaulttextstyle{\IMSauthorsize\@IEEEcompsocnotconfonly{\sffamily}\sublargesize}

\DeclareRobustCommand*{\IMSauthorrefmark}[1]{\raisebox{0pt}[0pt][0pt]{\textsuperscript{\footnotesize{#1}}}}%
\def\@IEEEauthorblockNtopspace{0ex}
\def\@IEEEauthorblockAtopspace{1mm}

\makeatother